# A homogeneous superconducting gap in DyBa$_2$Cu$_3$O$_{7-\delta}$ synthesized by oxide molecular beam epitaxy

Ze-Bin Wu[1*], Daniel Putzky[2], Asish K. Kundu[1], Hui Li[1,3], Shize Yang[1], Zengyi Du[1], Sang Hyun Joo[4], Jinho Lee[4], Yimei Zhu[1], Gennady Logvenov[2], Bernhard Keimer[2], Kazuhiro Fujita[1*], Tonica Valla[1*], Ivan Božović[1,5], and Ilya K. Drozdov[1*]

[1]Condensed Matter Physics and Materials Science Division, Brookhaven National Laboratory, Upton, NY 11973, USA
[2]Max Planck Institute for Solid State Research, Heisenbergstrasse 1, Stuttgart 70569, Germany
[3]Department of Physics and Astronomy, Stony Brook University, Stony Brook, NY 11794, USA
[4]Department of Physics and Astronomy, Seoul National University, Seoul 08826, Republic of Korea
[5]Department of Chemistry, Yale University, New Haven, CT 06520, USA.

*Corresponding to: zebinwu@bnl.gov, kfujita@bnl.gov, valla@bnl.gov, drozdov@bnl.gov.

# ABSTRACT

Much of what is known about high-temperature cuprate superconductors stems from studies based on two surface analytical tools, Angle-Resolved Photoemission Spectroscopy (ARPES) and Spectroscopic Imaging Scanning Tunneling Microscopy (SI-STM). A question of general interest is whether and when the surface properties probed by ARPES and SI-STM are representative of the intrinsic properties of bulk materials. We find this question is prominent in thin films of a rarely-studied cuprate, $DyBa_2Cu_3O_{7-\delta}$ (DBCO). We synthesize DBCO films by oxide molecular beam epitaxy and study them by *in-situ* ARPES and SI-STM. Both ARPES and SI-STM show that the surface DBCO layer is different from the bulk of the film — it is heavily underdoped, while the doping level in the bulk is very close to optimal doping evidenced by bulk-sensitive mutual inductance measurements. ARPES shows the typical electronic structure of a heavily underdoped $CuO_2$ plane and two sets of one-dimensional bands originating from the CuO chains with one of them gapped. SI-STM reveals two different energy scales in the local density of states, with one (at ~18 meV) corresponding to the superconductivity and the other one (at ~90 meV) to the pseudogap. While the pseudogap shows large variations over the length scale of a few nanometers, the superconducting gap is very homogeneous. This indicates that the pseudogap and superconductivity are of different origins.

## I. INTRODUCTION

Cuprates with a high superconducting transition temperature ($T_c$) are among the most intriguing strongly-correlated electronic systems [1]. They host a variety of complex electronic phases, such as d-wave superconductivity, strange metal, pseudogap, as well as charge, spin, and nematic ordering [1]. Cuprates usually also display various degrees of electronic inhomogeneity, which hampers clarification of the underlying physics. For example, the pseudogap has been attributed to preformed pairs or incoherent fluctuations of the pairing field (the so-called 'one-gap scenario') or alternatively to some of the competing phases (the 'two-gaps scenario'); this dichotomy is still unresolved [2-5]. We address it here by showing that in one representative high-$T_c$ cuprate superconductivity is remarkably homogeneous at low energy scales, in stark contrast to a heterogeneous pseudogap observed at higher energies, suggesting that high-$T_c$ superconductivity and pseudogap are of different origins.

Angle-Resolved Photoemission Spectroscopy (ARPES) and Spectroscopic Imaging Scanning Tunneling Microscopy (SI-STM) are among the most powerful techniques to study the electronic spectra of materials. ARPES features a high resolution in the momentum space, while SI-STM has the highest real-space resolution and can provide direct images of electronic (in)homogeneity. They are both extremely surface-sensitive with the probing depth on the 1 nm scale, which makes them powerful surface analytical tools but also brings about two problems. One is that the surface must be very clean and flat. In practice, this is usually achieved by cleaving bulk single crystals under ultrahigh vacuum (UHV). This, to a large extent, limits the choice of materials to those that cleave easily, such as $Bi_2Sr_2Ca_{n-1}Cu_nO_{2n+4}$ (BSCCO). The second issue is that high-$T_c$ superconductivity studies are typically focused on the intrinsic properties of bulk materials. Thus, the ultimate utility of SI-STM and ARPES data critically depends on whether the structure and electronic properties of the surface layer are the same as in the interior layers (or bulk). This is one of the key issues we answered in the present paper, on alert whether the answer is universal for the cuprates.

Together with BSCCO, hole-doped cuprates such as $La_{2-x}Ba_xCuO_4$ and $YBa_2Cu_3O_{7-\delta}$ (YBCO) have played an important role in the understanding of the various aspects of the high-$T_c$ superconductivity phenomenon. YBCO differs from the other cuprates by its unique structure in which CuO-chain layers supply carriers to the superconducting $CuO_2$ planes [6-8]. On cold-cleaved surfaces of YBCO crystals, ARPES resolved the electronic structure of $CuO_2$ planes similar to that in other hole-doped cuprates [9-14]. In addition to $CuO_2$ planes, the quasi-one-dimensional states corresponding to the CuO chains were also resolved. A recent study showed evidence for oxygen-deficient chains at the surface [15]. In these surface chains, a charge density wave (CDW) with the period corresponding to $2k_F$ (where $k_F$ is the Fermi wavevector) was observed in X-ray [16] and SI-STM studies [17-22], in addition to the CDW occurring in the $CuO_2$ plane [23,24].

While being the simplest and most common, cleaving is not the only method to prepare the samples of high-$T_c$ cuprates and other materials for surface-sensitive studies. Atomic layer-by-layer molecular beam epitaxy (ALL-MBE) is an advanced technique to synthesize complex materials with atomic precision [25-27] and is compatible with surface-sensitive analytic probes. To extend ARPES and SI-STM studies to other $RE$Ba$_2$Cu$_3$O$_{7-\delta}$ (i.e. $RE$-123, $RE$ denotes a trivalent rare-earth element) materials other than YBCO, and to avoid the problems of cleaving ionic materials, such as the polar catastrophe and multiple surface terminations, here we present the results on the synthesis of $DyBa_2Cu_3O_{7-\delta}$ (DBCO) films by ALL-MBE and the *in-situ* study by both ARPES and SI-STM of the same film. The ARPES study shows that the surface layer of DBCO film is heavily underdoped because of a substantial loss of oxygen. This is contrary to many cases in cleaved YBCO, where the surface component of the ARPES spectrum is heavily overdoped regardless of the oxygen stoichiometry of the bulk crystal [12,13]. In either case, the surface layers are substantively different from the bulk regarding to the hole doping content. The contrast between the surfaces of the cleaved YBCO bulk crystal and the DBCO films underscores the nonnegligible effects of material preparation, particularly relevant for surface-sensitive probes.

## II. EXPERIMENT

The DBCO film growth and the key characterizations were conducted in the OASIS facility [28], which integrates oxide MBE, ARPES, and SI-STM. The base pressure of the MBE chamber is $8 \times 10^{-10}$ Torr. Essentially 100%-pure liquid ozone is generated and stored in a cryogenic still and is introduced into the MBE chamber through a gas injector pointing at the sample, the end of which is close to the sample so that the pressure of ozone is about 100 mTorr at the nozzle of the injector. The background ozone pressure during growth is set to $3 \times 10^{-5}$ Torr. The sample is heated by an infra-red bulb from the back and the temperature is read using an optical pyrometer from the front. Before the synthesis, the calibration of each source is performed in the ozone atmosphere by a quartz-crystal monitor, which is placed at the same position where the sample is kept during synthesis. The substrate is 0.7% Nb-doped $SrTiO_3$ polished with the surface perpendicular to the crystallographic [001] direction. It has a high absorption coefficient for infrared radiation, avoiding the back-coating process required for transparent substrates.

The as-grown samples were transferred to ARPES and SI-STM under UHV. The ARPES measurements were carried out using a Scienta SES-R4000 electron spectrometer with the monochromatized He-IIα (40.8 eV) radiation (VUV-5k). The total instrumental energy resolution is around 20 meV for the ARPES measurements. Angular resolution is better than 0.15° and 0.3° along and perpendicular to the slit of the analyzer, respectively. Most of the data were taken at 35 K, except for the room temperature (T = 300 K) measurements. The SI-STM measurements were carried out at 9 K. The topography was measured in the constant-current mode typically at 3 GΩ junction resistance ($V_{bias}$ = 150 meV and $I_t$ = 50 pA). The differential conductance maps g(r, E) were obtained by the standard lock-in technique at 875.5 Hz with 4 mV bias modulation on a 200×200 pixels grid within 150meV energy range with a spectroscopic setpoint typically at the same 3 GΩ junction resistance.

After in-situ characterizations in ARPES and SI-STM, $T_c$ of the sample was measured ex situ in mutual inductance [29]. The sample was cooled down continuously from the room temperature to T ~ 4 K. The inductance coils have a large number of turns (300 for driving coil and 900 for pick-up coil, respectively) but a very small inner radius (250 $\mu m$ for driving coil and 500 $\mu m$ for pick-up coil), much smaller than the film size (5 × 10 mm$^2$), so that the field leakage around the film is minimal. The frequency of the driving current is 40 kHz. In superconducting films, the HWHM ($\delta T$) of the peak in ImM(T), can be taken as a measure (an upper bound) to the spread of $T_c$ in that film. For example, if the film contains two large lateral domains with $T_{c1}$ and $T_{c2}$, respectively, that differed by more than this $\delta T$, one would see two resolved peaks. If there were a continuous (or a discrete but dense) spread of $T_c$ in the interval $T_1 < T_c < T_2$, one would see one peak in ImM(T) but broader than $T_2 - T_1$.

After the measurement of $T_c$, the cross-section profile of the film was imaged by scanning transmission electron microscopy (STEM) with the JEOL ARM200CF electron microscope which is equipped with two aberration correctors, a cold-field electron source and a direct electron detector.

## III. RESULTS

### A. Synthesis of DBCO films by MBE

The crystal structure of DBCO is shown in Fig. 1a. The parent antiferromagnetic compound with $\delta = 1$ has a tetragonal structure without oxygen in the chain layers, whereas the Cu $d_{x^2-y^2}$ orbitals in the $CuO_2$ planes are occupied with exactly one electron, resulting in a charge-transfer insulator. As O(1) vacancies are filled, CuO chains build up gradually, the structure becomes orthorhombic, the hole content ($p$) in $CuO_2$ planes increases, and at $p \approx 0.05$ superconductivity emerges [30,31].

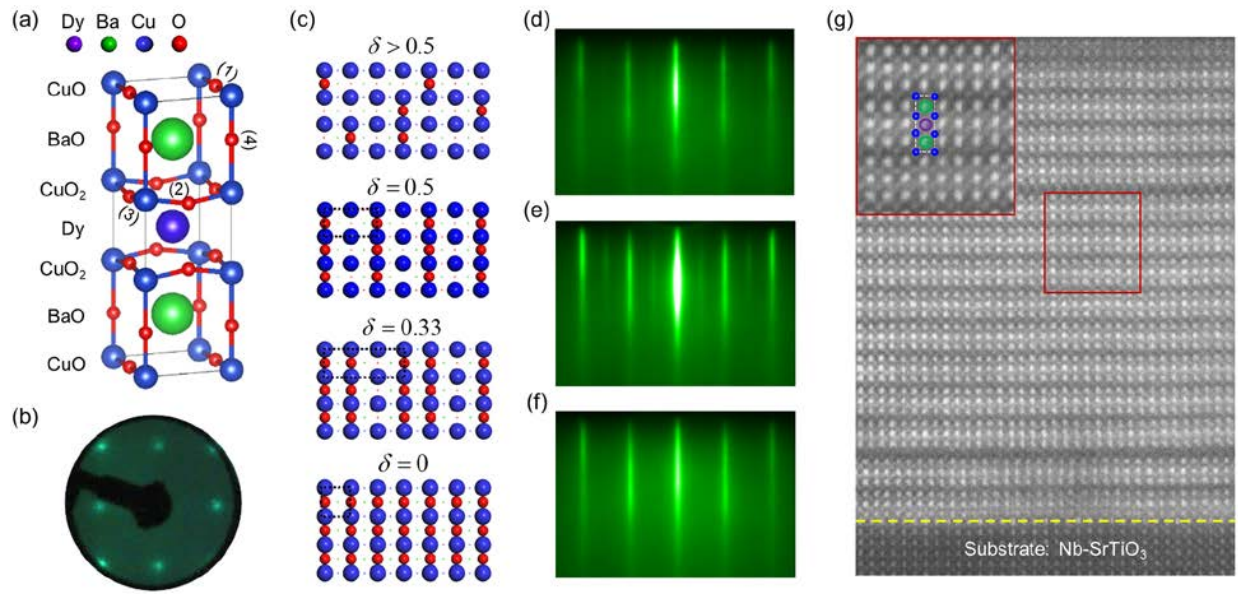

**Fig. 1**. Structure of a $DyBa_2Cu_3O_{7-\delta}$ (DBCO) film. **a** The unit cell of DBCO ($\delta = 0$ in this model). **b** A LEED pattern of the DBCO film after the synthesis, taken at electron beam energy of 67 eV. **c** Atomic models of CuO chains ordering with different oxygen deficiency $\delta$ in the CuO-chain plane. (d)-(f) RHEED patterns captured at electron beam energy of 30 keV: **d** after the synthesis before cooling down; **e** in the middle of cooling down around 400 °C; **f** after cooling down and with ozone shut off. **g** A cross-section profile of the DBCO film by aberration corrected STEM. The inset shows the zoomed-in image of the area marked with the red rectangle, with the unit cell model placed on top as a guide for atom identification. The interface between the Nb-doped $SrTiO_3$ substrate and the DBCO film is marked by the yellow dashed line.

The difficulty of synthesizing DBCO by ALL-MBE originates in the existence of many competing impurity phases. The appearance of unwanted secondary-phase precipitates depends strongly on the growth temperature and can be triggered by even slight off-stoichiometry of the elements.

Therefore, the optimization of the growth parameters and real-time control of the shuttering of atomic-beam sources are crucially important. Once the optimal conditions are established, the atomic layer-by-layer growth of DBCO is manifested by the periodic intensity oscillation of reflection high-energy electron diffraction (RHEED) patterns, matching the periodic growth for each unit cell. The RHEED pattern of a 20-unit-cell-thick film recorded right after the synthesis at the growth temperature > 600°C (Fig. 1d) shows sharp diffraction rods. The distance between the main streaks corresponds to the in-plane lattice periodicity (~ 0.39 nm) of epitaxial DBCO films on $SrTiO_3$. The atomic layer-by-layer growth is confirmed by the cross-section profile of this film (Fig. 1g) characterized *ex situ* by STEM, in which the atomic structure is clearly resolved and matches well with the atomic model (Fig. 1a) except for few stacking faults.

The same film was measured *ex situ* by mutual inductance [29] and the results are shown in Fig. 2. The imaginary part of mutual inductance, Im$M$(T), shows a peak that upon cooling onsets at $T_c$ ≈ 79 K, the temperature at which the DC resistivity has dropped to zero. The transition is very narrow, giving rise to a sharp peak in Im$V_p$ (Fig. 2b). The corresponding half-width-at-half-maximum (HWHM) of that peak, $\delta T$ ≈ 0.5 K, can be taken as a measure (an upper bound) to the lateral spread of $T_c$ in that film. In the DBCO film under study, $T_c$ is homogeneous at the level of ± 0.3 % over the length scale of several millimeters.

According to the early studies of different members of the *RE*-123 family of materials, $T_c$ depends very little on the choice of the specific rare-earth element [32] and a comparison to the well-studied phase diagram of YBCO is possible. Accordingly, the present film is underdoped, but *p* is not far from the nominal optimal doping level of $p$ ≈ 0.16, especially when one takes into account the suppressing effect of epitaxial strain on superconductivity of DBCO films [33].

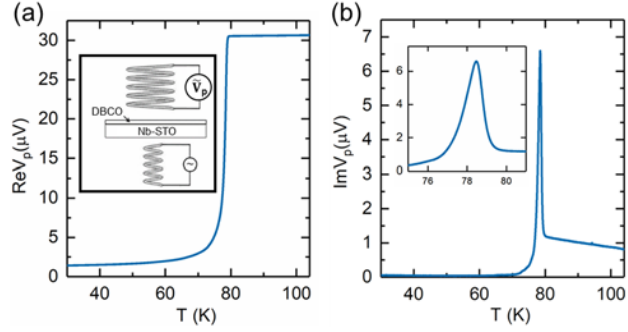

**Fig. 2.** Superconductivity and determination of $T_c$ of the DBCO film by mutual-inductance measurements. a The real component of pick-up signal $V_p$, the voltage across the pickup coil, showing diamagnetic screening (the Meissner effect) when the film becomes superconducting. Inset is the schematic of the experimental set-up, where the driving coil is under the substrate and the pick-up coil is above the film. b The imaginary part of $V_p$, indicating that $T_c \approx 79$ K, with the transition HWHM of ~0.5 K. The metallic Nb-STO substrate has a slight diamagnetic screening effect, which causes the slow ramp of $ImV_p$ above the superconducting phase transition of the DBCO film.

To assess the surface quality before the ARPES and SI-STM studies, the same film was characterized by low-energy electron diffraction (LEED) at room temperature (Fig. 1b). The pattern of the diffraction spots confirms the quasi-square lattice at the surface. Some surface reconstructions were detected by RHEED during the cooling down of the film in ozone atmosphere. Figure 1e is the pattern captured at an intermediate temperature, around 400°C, where the intensity of streaks in the middle of the main streaks gets enhanced, indicating a strong 2-fold reconstruction of the surface structure. This characteristic pattern persists for a while as the temperature is decreased further and then disappears at around 300°C, with only the main streaks remaining after cooling down (Fig. 1f). According to the growth sequence [33] (2Ba-Dy-3Cu), the top Cu-O layer of the DBCO film should be a chain layer. Then the surface reconstruction can be related to the CuO chains ordering, which evolves with the oxygen deficiency [34,35]. Several illustrative atomic models of the dependence on $\delta$ of CuO chains ordering are presented in Fig. 1c. After the last Cu deposition and before cooling down, $\delta$ at the surface is supposed to be at the same level as that in the bulk layers, which is < 0.5. The enhancement of the 2-fold side streaks in the RHEED pattern during cooling down can be related to the chains model with $\delta = 0.5$, considering that the surface reconstruction can be caused by the oxygen loss during the cool down process.

## B. In-situ study of DBCO film by ARPES

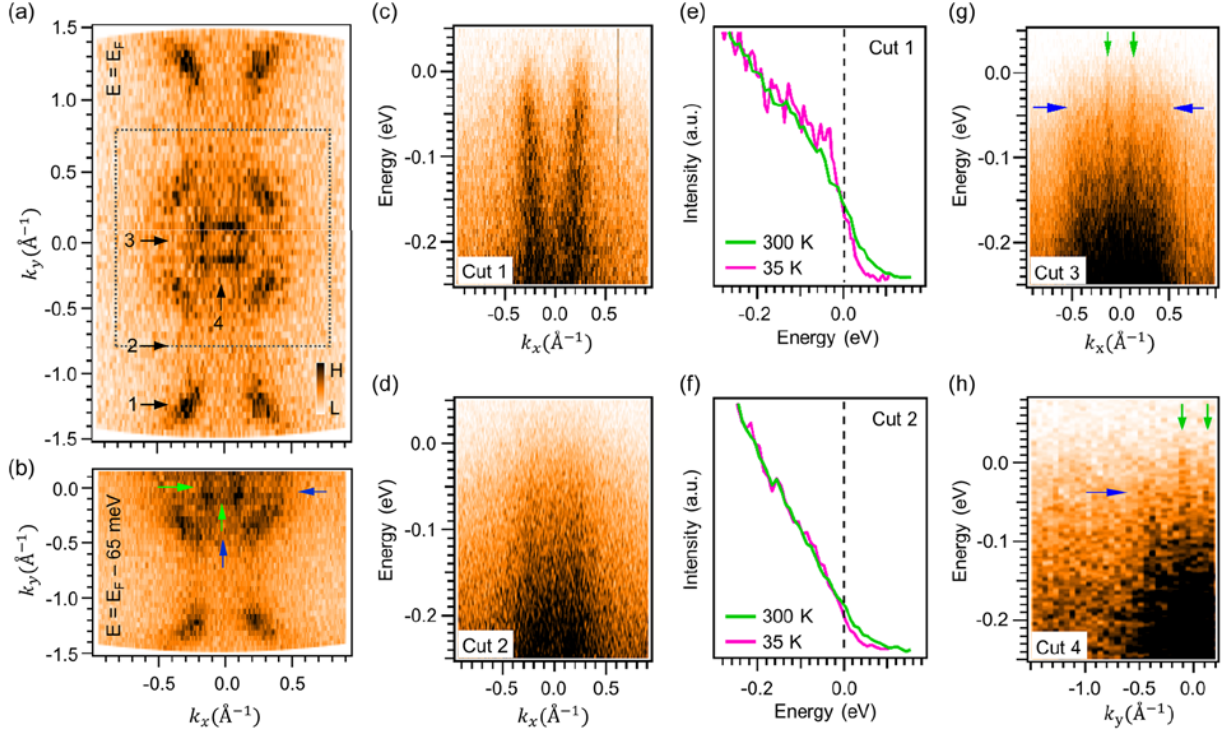

**Fig. 3.** Fermi surface and band dispersions of the DBCO film. a Fermi surface map showing both $CuO_2$-plane and CuO-chain features (The map is symmetrized with respect to $k_y = 0$). The first Brillouin zone is marked with the dotted square. The intensity bar shown at the right bottom applies to all panels. b Iso-energy surface at $E = E_F - 65$ meV, showing the two sets of CuO-chain states originating from the subsurface (green arrows) and the surface (blue arrows). c-d Energy-momentum dispersions along the momentum cuts 1 and 2 in a, respectively, taken at $T = 35$ K. Fermi level is set to 0 eV for c-h. e-f Energy distribution curves (EDCs) corresponding to the cuts 1 and 2, below and above $T_c$ as indicated. EDCs corresponding to cut 2 shows suppression of intensity near the Fermi level due to the pseudogap of $CuO_2$ planes. g-h Energy-momentum dispersions along momentum cuts 3 and 4, respectively, showing dispersions of two sets of orthogonal CuO-chain bands. The set of chains states indicated by blue arrows shows a gap-like feature with the suppression of intensity near the Fermi level.

In Figure 3a, we show the photoemission intensity at the Fermi level for the same DBCO film as discussed in the previous section. The states originating from $CuO_2$ planes and CuO chains are clearly distinguishable. From the size and the shape of the Fermi contour, it is apparent that the $CuO_2$ planes probed by ARPES are heavily underdoped [12,13], less than 0.1. In addition to the states corresponding to the $CuO_2$ planes, the two sets of CuO-chain states perpendicular to each other can be identified near the zone center, similar to cleaved YBCO [13]. This indicates that the CuO chains in *RE*-123 align to either [100] or [010] substrate directions because of twinning.

The energy-momentum dispersion spectra along cut 1 (Fig. 3c), which traverses the nodal region, shows a dispersion characteristic of the $CuO_2$ plane with no gap at the Fermi energy. The spectrum along cut 2 (Fig. 3d), which traverses the antinodal region, shows a suppressed intensity near the Fermi energy due to the pseudogap (~ 60 meV) affecting the $CuO_2$ planes. In the spectra taken along $k_y = 0$ (cut 3) and $k_x = 0$ (cut 4), as shown in Fig. 3g and 3h, there are two sets of CuO-chain states with very different dispersions. One set is metallic, crossing the Fermi level at $|k_F| \sim 0.13$ Å$^{-1}$, while the other one is gapped (~ 35 meV), and with a much larger $|k_F| \sim 0.44$ Å$^{-1}$.

In the intensity map at $E = E_F - 65$ meV (Fig. 3b), both sets of CuO-chain bands are visible. By comparing with the data[14] from cleaved YBCO, the bands marked by the green arrows in Fig. 3b, 3g, and 3h are from the subsurface, while those marked by the blue arrows are from the surface. The gap in the surface chains originates from the loss of oxygen, and according to a previous ARPES study [14] on YBCO could be an indication of the formation of CDW at the surface. Based on the assumption that the CDW is due to a $2k_F$ instability of surface chains, the period of CDW would be ~7 Å, while at the CuO chains $\delta$ should be close to 0.5. It is not surprising that such a large concentration of oxygen vacancies (charged defects) introduces a high level of surface disorder, as well as making the probed $CuO_2$ planes heavily underdoped.

The energy distribution curves (EDCs) obtained from both cut 1 and cut 2 are presented in Fig. 3e and 3f. The shift of the leading edge of EDC from the antinodal region of $CuO_2$ planes relative to that from the nodal region indicates a pseudogap affecting that region of the Brillouin zone. However, there is no additional measurable shift (> 5 meV) when the sample is cooled down from 300 K to 35 K along the cut 2, indicating that superconductivity in the film has a negligible effect on the spectral properties of $CuO_2$ planes in the antinodal region at the surface.

## C. In-situ study of DBCO film by SI-STM

Figure 4a shows the STM topography of the DBCO film measured at 9 K within a $20 \times 20$ nm$^2$ field of view, in which randomly formed atomic clusters (about 20 Å wide) are observed. The one-dimensional CuO chain-like structures that were previously reported [17-22] for cleaved YBCO are not apparent here. This can be caused by the oxygen vacancies and related surface reconstructions due to oxygen loss at the surface of the film, as indicated by both RHEED and ARPES.

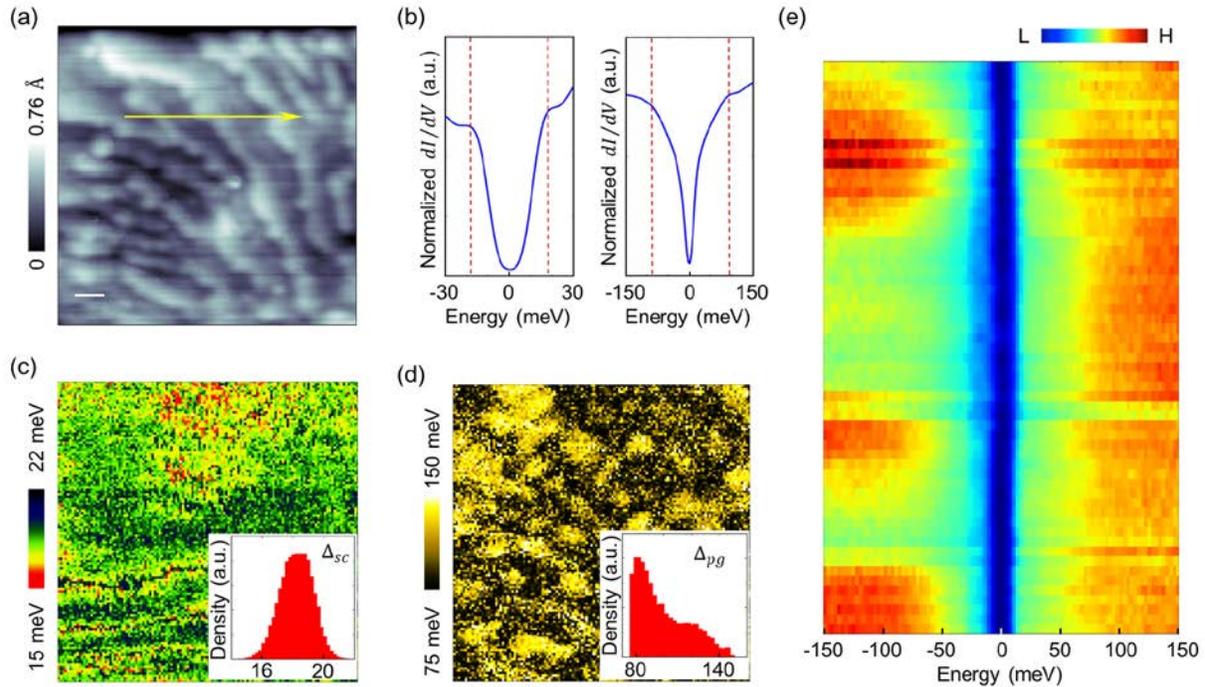

**Fig. 4.** Spectroscopic imaging of the DBCO film. a Topography of a $20 \times 20$ nm2 field of view, where a differential conductance spectrum is measured within ±150 meV at each location on a $200 \times 200$ pixels grid. b Spatially-averaged LDOS spectra for both low (|E| < 30 meV) and high energies (|E |< 150 meV). c-d Spatial variations of the gaps at low-energy and high-energy scales, with insets showing the corresponding distribution histograms. e LDOS spectra taken along the line cut in a, showing spatially uniform low-energy states and heterogeneous high-energy states.

In the same field of view, the differential conductance spectrum, which is proportional to the local density of states (LDOS), is measured within ±150 meV at each location and mapped out on a 200 × 200 pixels grid. In Fig. 4b, spatially averaged LDOS spectra for both low ($|E| < 30$ meV) and high energies ($|E| < 150$ meV) are plotted, exhibiting a particle-hole symmetric gap below $|E| < 30$ meV without additional states arising from the chains. Although no well-defined peaks are observed in the LDOS spectra, there are obviously two different energy scales – one is a shoulder around ± 18 meV with a U-shaped-like spectrum that deviates from the pure d-wave scenario, and the other one is a broad hump around ± 90 meV, as indicated by broken lines in Fig. 4b.

Empirically, one can define a gap ($2\Delta$) as the energy separation between two coherence peaks. In Fig. 4c and 4d, spatial variations of the low ($\Delta_0$) and high ($\Delta_1$) energy scales are presented, respectively, with insets showing their distributions. It is worth noting that the $\Delta_0(\boldsymbol{r})$ is highly homogeneous with a spatial average of ~18 meV and a spatial variation of ±2 meV with respect to the average value. Such a variation (the full-width-at-half-maximum ~ 10%) is in fact much smaller than those reported in heavily overdoped $Bi_2Sr_2CaCu_2O_{8+\delta}$ (Bi2212) (~ 40%), which is believed to be relatively homogeneous [36]. In contrast, $\Delta_1(\boldsymbol{r})$ is highly heterogeneous with ± 30 meV variations from the spatial average of ~ 90 meV. Such contrasting behavior is apparent in the line cut of the LDOS spectra shown in Fig. 4e. While the low-energy states near $\Delta_0$ are spatially uniform as indicated by uniform blue color, the high-energy states close to $\Delta_1$ manifest strong electronic heterogeneity.

In general, these observations are consistent with those reported on underdoped Bi2212 [36-38], in which the energy scale on which the Bogoliubov quasiparticle interference and the kink in spec-tra appear is pretty uniform at low energies. In contrast, the pseudogap energy is quite heterogeneous. If we assume that $T_c$ and the superconducting gap $\Delta_0$ are connected according to the phenomenological formula $2\Delta_0/kT_c \sim 7$ obtained from YBCO bulk crystal [39], then $T_c = 79$ K should indicate a superconducting gap of ~ 25 meV. Considering that the surface is underdoped as evidenced by ARPES, the superconducting gap probed by SI-STM is expected to be lower, which is consistent with the low energy scale of $\Delta_0 \sim 18$ meV. Thus, we infer from these observations that $\Delta_0$ is related to the superconducting gap, while $\Delta_1$ corresponds to the pseudogap.

## IV. DISCUSSION AND OUTLOOK

In summary, using ALL-MBE we have successfully synthesized high-quality DBCO thin films and studied them *in situ* by both ARPES and SI-STM. Both the chain and the plane electronic structures of DBCO have been clearly resolved in ARPES and are consistent with the previous observations on YBCO single crystals. SI-STM has shown features at ~18 meV and ~90 meV attributed to the superconducting gap and the pseudogap, respectively. The data allow us to derive conclusions pertinent to the two key questions we have posited in the Introduction — whether the surface layer is the same as the interior layers (or bulk), and whether the pseudogap is a precursor of the superconductivity or not.

According to our RHEED, the DBCO surface is unstable, resulting in a surface reconstruction most likely related to the loss of oxygen. Our ARPES data, in particular the shape and the size of the Fermi surface, and the observation of the gap in the chain electronic structure, show that the surface is heavily underdoped. The SI-STM data are also consistent with the underdoped surface layer; the observed superconducting gap is smaller while the pseudogap is larger than what is expected based on the bulk $T_c = 79$ K. Therefore, we conclude that in DBCO films synthesized by MBE the surface layer is indeed very different from the bulk. Specifically, due to the substantial loss of oxygen, the surface layer is heavily underdoped relative to the bulk of the same film. In contrast, the cleaved surface of YBCO bulk crystal was found to be heavily overdoped in many cases [12,13]. In either case, in *RE*-123 materials the surface is different from the bulk (layers), while the doping level of $CuO_2$ planes near the surface varies depending on the preparation method.

On the other hand, the cohesion energy in *RE*-123 is primarily ionic. Madelung interactions are long-ranged in ionic materials and any given layer is not independent of the other layers [40,41]. Since the surface layer exists in a different effective environment than the bulk, its structure is expected to change. It has been demonstrated that Madelung (electrostatic) strain propagates from the substrate through 40 unit cells thick LSCO layers, modifying the *c*-axis lattice constant of the subsequent layers [40]. A similar effect was also observed [41] in YBCO. Synchrotron-based, coherent X-ray diffraction experiments have demonstrated a very large displacement ($> 0.5$ Å) of the apical oxygen in the top layer in LSCO-based heterostructures [42]. While to what extent such structural changes in ionic-type cuprates would modify the electronic properties at the surface, especially probed by ARPES and SI-STM, still calls for further studies for specific materials.

However, the situation in BSCCO is rather different, because it is a naturally layered material with weak (van der Waals) bonds between layers, which makes the layers basically independent from one another. Indeed, studies on single-monolayer BSCCO samples and bulk BSCCO by STM and transport measurements [43] revealed that the transport measurements (that probe the entire volume) agree with the results from SI-STM and ARPES (that probe the surface only).

The answer to our question, then, is that in some cuprates, like *RE*-123, the top layer is indeed different from the bulk, showing changes in the stoichiometry (such as the oxygen content) as well as various forms of structural reconstructions for films synthesized by MBE. In BSCCO, this is clearly not the case due to the weak interlayer van der Waals bonding [43]. This is an important validation for the HTS community because the knowledge about the electronic spectra of cuprates has indeed been predominantly based on ARPES and SI-STM studies of BSCCO.

We now turn to the question of how are the pseudogap and the superconducting gap related. Our SI-STM data indicate that there is a qualitative difference: the pseudogap is very inhomogeneous (to about ± 30 meV) on the (lateral) length scale of a few nanometers, while the superconducting gap is uniform (to about ± 2 meV). The magnetic mutual inductance measurements show that the superconductivity is indeed also very homogeneous on a macroscopic scale in the bulk. These qualitatively different behaviors of the superconducting gap and the pseudogap seem to indicate that they are of different origins and possibly competing. Ultimately, it is a question for theory whether this interpretation is unique.

From a technical viewpoint, the present study proves the unique value of the new facility ("OASIS") [28] at Brookhaven National Laboratory, which incorporates the ALL-MBE synthesis with ARPES and SI-STM modules, all interconnected under UHV. OASIS was designed and built to enable answering systematically the type of questions considered here. ALL-MBE enables sample engineering at a single atomic layer level, so in the future, we anticipate synthesizing and studying one-unit-cell-thick layers of various HTS cuprates and other materials of interest on 'neutral' substrates or buffers, and comparing their ARPES and SI-STM spectra to those taken from bulk samples (or thicker films) of the same material. Alternatively, ALL-MBE also enables the synthesis of ultrathin layers of a given material on 'active' buffers or substrates, and the study of proximity effects in heterostructures. This may be of great interest to, e.g., the quest for and study of topological superconductors.

**ACKNOWLEDGEMENTS**

The authors thank A. N. Pasupathy and P. D. Johnson for useful discussions. This work was supported by the US Department of Energy, Office of Basic Energy Sciences, contract no. DE-SC0012704. Z.B.W., A.K.K., H.L., Z.Y.D., K.F., T.V., Y.Z., I.B., and I.K.D. acknowledge support from the U.S. Department of Energy, Office of Basic Energy Sciences, under contract number DEAC02-98CH10886. D.P., G.L., and B.K. acknowledge funding from the Deutsche Forschungsgemeinschaft (DFG, German Research Foundation), Projekt No. 107745057 TRR 80 and from the European Union's Horizon 2020 research and innovation programme under Grant Agreement No. 823717ESTEEM3. S.H.J. and J.L. acknowledge support from the Institute for Basic Science in Korea (Grant No. IBS-R009-G2), the Institute of Applied Physics of Seoul National University, and National Research Foundation of Korea (NRF) grant funded by the Korea government (MSIP) (No. 2017R1A2B3009576).

# REFERENCES

[1] B. Keimer, S. A. Kivelson, M. R. Norman, S. Uchida, and J. Zaanen, From quantum matter to high-temperature superconductivity in copper oxides, Nature **518**, 179 (2015).

[2] V. J. Emery and S. A. Kivelson, Importance of phase fluctuations in superconductors with small superfluid density, Nature **374**, 434-437 (1994).

[3] S. Hüfner, M. A. Hossain, A. Damascelli, and G. A. Sawatzky, Two gaps make a high-temperature superconductor? Rep. Prog. Phys. **71**, 062501 (2008).

[4] T. Yoshida, W. Malaeb, S. Ideta, D. H. Lu, R. G. Moor, Z.-X. Shen, M. Okawa, T. Kiss, K. Ishizaka, S. Shin, Seiki Komiya, Yoichi Ando, H. Eisaki, S. Uchida, and A. Fujimori, Coexistence of a pseudogap and a superconducting gap for the high-$T_c$ superconductor $La_{2-x}Sr_xCuO_4$ studied by angle-resolved photoemission spectroscopy, Phys. Rev. B **93**, 014513 (2016).

[5] I. Božović and J. Levy, Pre-formed Cooper pairs in copper oxides and $LaAlO_3/SrTiO_3$ heterostructures, Nature Phys. **16**, 712-717 (2020).

[6] J. Zaanen, A. T. Paxton, O. Jepsen, and O. K. Anderson, Chain-fragment doping and the phase diagram of $YBa_2Cu_3O_{7-\delta}$, Phys. Rev. Lett. **60**, 2685 (1988).

[7] R. J. Cava, A. W. Hewat, E. A. Hewat, B. Batlogg, M. Marezio, K. M. Rabe, J. J. Krajewski, W. F. Peck Jr, and L. W. Rupp Jr, Structural anomalies, oxygen ordering and superconductivity in oxygen deficient $Ba_2YCu_3O_x$, Physica C: Superconductivity **165**, 419 (1990).

[8] J. L. Tallon, C. Bernhard, H. Shaked, R. L. Hitterman, and J. D. Jorgensen, Generic superconducting phase behavior in high-$T_c$ cuprates: $T_c$ variation with hole concentration in $YBa_2Cu_3O_{7-\delta}$, Phys. Rev. B **51**, 12911 (1995).

[9] M. C. Schabel, C.-H. Park, A. Matsuura, Z,-X, Shen, D. A. Bonn, R. Liang, and W. N. Hardy, Angle-resolved photoemission on untwinned $YBa_2Cu_3O_{6.95}$, I. electronic structure and dispersion relations of surface and bulk bands. Phys. Rev. B **57**, 6090 (1998).

[10] D. H. Lu, D. L. Feng, N. P. Armitage, K. M. Shen, A. Damascelli, C. Kim, F. Ronning, Z.-X. Shen, D. A. Bonn, R. Liang, W. N. Hardy, A. I. Rykov, and S. Tajima, Superconducting gap and strong in-plane anisotropy in untwinned $YBa_2Cu_3O_{7-\delta}$, Phys. Rev. Lett. **86**, 4370 (2001).

[11] S.V. Borisenko, A. A. Kordyuk, V. Zabolotnyy, J. Geck, D. Inosov, A. Koitzsch, J. Fink, M. Knupfer, B. Büchner, V. Hinkov, C. T. Lin, B. Keimer, T. Wolf, S. G. Chiuzbaian, L. Patthey, and R. Follath, Kinks, nodal bilayer splitting, and interband scattering in $YBa_2Cu_3O_{6+x}$, Phys. Rev. Lett. **96**, 117004 (2006).

[12] V. B. Zabolotnyy, S. V. Borisenko, A. A. Kordyuk, J. Geck, D. S. Inosov, A. Koitzsch, J. Fink, M. Knupfer, B. Büchner, S.-L. Drechsler, H. Berger, A. Erb, M. Lambacher, L. Patthey, V. Hinkov, and

B. Keimer, Momentum and temperature dependence of renormalization effects in the high-temperature superconductor YBa$_2$Cu$_3$O$_{7-\delta}$, Phys. Rev. B **76**, 064519 (2007).

[13] M. A. Hossain, J. D. F. Mottershead, D. Fournier, A. Bostwick, J. L. McChesney, E. Rotenberg, R. Liang, W. N. Hardy, G. A. Sawatzky, I. S. Elfimov, D. A. Bonn, and A. Damascelli, In situ doping control of the surface of high-temperature superconductors, Nature Phys. **4**, 527 (2008).

[14] V. B. Zabolotnyy, A. A. Kordyuk, D. Evtushinsky, V. N. Strocov, L. Patthey, T. Schmitt, D. Haug, C. T. Lin, V. Hinkov, B. Keimer, B. Büchner, and S. V. Borisenko, Pseudogap in the chain states of YBa$_2$Cu$_3$O$_{6.6}$, Phys. Rev. B **85**, 064507 (2012).

[15] H. Iwasawa, P. Dudin, K. Inui, T. Masui, T. K. Kim, C. Cacho, and M. Hoesch, Buried double CuO chains in YBa$_2$Cu$_4$O$_8$ uncovered by nano-ARPES, Phys. Rev. B **99**, 140510 (2019).

[16] A. J. Achkar, R. Sutarto, X. Mao, F. He, A. Frano, S. Blanco-Canosa, M. Le Tacon, G. Ghiringhelli, L. Braicovich, M. Minola, M. Moretti Sala, C. Mazzoli, Ruixing Liang, D. A. Bonn, W. N. Hardy, B. Keimer, G. A. Sawatzky, and D. G. Hawthorn, Distinct charge orders in the planes and chains of ortho-III-ordered YBa$_2$Cu$_3$O$_{6+\delta}$ superconductors identified by resonant elastic x-ray scattering. Phys. Rev. Lett. **109**, 167001 (2012).

[17] H. L. Edwards, A. L. Barr, J. T. Markert, and A. L. de Lozanne, Modulations in the CuO chain layer of YBa$_2$Cu$_3$O$_{7-\delta}$: charge density waves? Phys. Rev. Lett. **73**, 1154 (1994).

[18] H. L. Edwards, D. J. Derro, A. L. Barr, J. T. Markert, and A. L. de Lozanne, Spatially varying energy gap in the CuO chains of YBa$_2$Cu$_3$O$_{7-\delta}$ detected by scanning tunneling spectroscopy, Phys. Rev. Lett. **75**, 1387 (1995).

[19] D. J. Derro, E.W. Hudson, K.M. Lang, S. H. Pan, J. C. Davis, J. T. Markert, and A. L. de Lozanne, Nanoscale one-dimensional scattering resonances in the CuO chains of YBa$_2$Cu$_3$O$_{6+x}$, Phys. Rev. Lett. **88**, 097002 (2002).

[20] M. Maki, T. Nishizaki, K. Shibata, and N. Kobayashi, Electronic structure of the CuO-chain layer in YBa$_2$Cu$_3$O$_{7-\delta}$ studied by scanning tunneling microscopy, Phys. Rev. B **65**, 140511 (2002).

[21] M. Maki, T. Nishizaki, K. Shibata, and N. Kobayashi, Layered charge-density waves with nanoscale coherence in YBa$_2$Cu$_3$O$_{7-\delta}$, Phys. Rev. B **72**, 024536 (2005).

[22] M. Maki, T. Nishizaki, K. Shibata, and N. Kobayashi, Low-temperature scanning tunneling microscopy of YBa$_2$Cu$_3$O$_{7-\delta}$, Physica C: Superconductivity **378-381**, 84 (2002).

[23] T. Wu, H. Mayaffre, S. Krämer, M. Horvatić, C. Berthier, W. N. Hardy, R. Liang, D. A. Bonn, and M.-H. Julien, Magnetic-field-induced charge-stripe order in the high-temperature superconductor YBa$_2$Cu$_3$O$_y$, Nature **477**, 191 (2011).

[24] J. Chang, E. Blackburn, A. T. Holmes, N. B. Christensen, J. Larsen, J. Mesot, R. Liang, D. A. Bonn, W. N. Hardy, A.Watenphul, M. v. Zimmermann, E. M. Forgan, and S. M. Hayden, Direct observation

of competition between superconductivity and charge density wave order in YBa$_2$Cu$_3$O$_{6.67}$, Nature Phys. **8**, 871 (2012).

[25] A. Gozar, G. Logvenov, L. Fitting Kourkoutis, A. T. Bollinger, L. A. Giannuzzi, D. A. Muller, and I. Božović, High-temperature interface superconductivity between metallic and insulating copper oxides, Nature **455**, 782 (2008).

[26] I. Božović, X. He, J. Wu, and A. T. Bollinger, Dependence of the critical temperature in overdoped copper oxides on superfluid density, Nature **536**, 309 (2016).

[27] G. Logvenov, A. Gozar, and I. Božović, High-temperature superconductivity in a single copper-oxygen plane, Science **326**, 699 (2009).

[28] C. K. Kim, I. K. Drozdov, K. Fujita, J. C. Séamus Davis, I. Božović, T. Valla, In-situ angle-resolved photoemission spectroscopy of copper-oxide thin films synthesized by molecular beam epitaxy, J. Electron. Spectrosc. Relat. Phenom. (2018).

[29] X. He, A. Gozar, R. Sundling, and I. Božović, High-precision measurement of magnetic penetration depth in superconducting films, Rev. Sci. Instrum. 87, 113903 (2016).

[30] R. Liang, D. A. Bonn, and W. N. Hardy, Evaluation of CuO$_2$ plane hole doping in YBa$_2$Cu$_3$O$_{6+x}$ single crystals, Phys. Rev. B **73**, 180505 (2006).

[31] M. R. Presland, J. L. Tallon, R. G. Buckley, R. S. Liu, and N. E. Flower, General trends in oxygen stoichiometry effects on $T_c$ in Bi and Tl superconductors, Physica C: Superconductivity 176, 95 (1991).

[32] K. N. Yang, Y. Dalichaouch, J. M Ferreira, B. W. Lee, J. J. Neumeier, M. S. Torikachvili, H. Zhou, and M. B. Maple, High temperature superconductivity in rare-earth (R)-barium copper oxides (RBa$_2$)Cu$_3$O$_{9-\delta}$, Solid State Commun. **63**, 515 (1987).

[33] D. Putzky, P. Radhakrishnan, Y. Wang, P. Wochner, G. Christiani, M. Minola, P. A. van Aken, G. Logvenov, E. Benckiser, and B. Keimer, Strain-induced structural transition in DyBa$_2$Cu$_3$O$_{7-x}$ films grown by atomic layer-by-layer molecular beam epitaxy, Appl. Phys. Lett. **117**, 072601 (2020).

[34] N. H. Andersen, M. von Zimmermann, T. Frello, M. Käll, D. Mønster, P.-A. Lindgard, J. Madsen, T. Niemöller, H. F. Poulsen, O. Schmidt, J. R. Schneider, Th. Wolf, P. Dosanjh, R. Liang, and W. N. Hardy, Superstructure formation and the structural phase diagram of YBa$_2$Cu$_3$O$_{6+x}$, Physica C: Superconductivity **318**, 259 (1999).

[35] J. R. Schneider, M. von Zimmermann, T. Frello, N. H. Andersen, J. Madsen, M. Käll, H. F. Poulsen, R. Liang, P. Dosanjh, and W. N. Hardy, Oxygen-ordering superstructures in underdoped YBa$_2$Cu$_3$O$_{6+x}$ studied by hard x-ray diffraction, Phys. Rev. B **68**, 104515 (2003).

[36] J. W. Alldredge, J. Lee, K. McElroy, M. Meng, K. Fujita, Y. Kohsaka, C. Taylor, H. Eisaki, S. Uchida, P. J. Hirschfeld, and J. C. Davis, Evolution of the electronic excitation spectrum with strongly diminishing hole density in superconducting $Bi_2Sr_2CaCu_2O_{8+\delta}$. Nature Phys, **4**, 319 (2008).

[37] K. McElroy, D.-H. Lee, J. E. Hoffman, K. M. Lang, J. Lee, E.W. Hudson, H. Eisaki, S. Uchida, and J. C. Davis, Coincidence of checkerboard charge order and antinodal state decoherence in strongly underdoped superconducting $Bi_2Sr_2CaCu_2O_{8+\delta}$, Phys. Rev. Lett. **94**, 197005 (2005).

[38] K. Tanaka, W. S. Lee, D. H. Lu, A. Fujimori, T. Fujii, Risdiana, I. Terasaki, D. J. Scalapino, T. P. Devereaux, Z. Hussain, Z.-X. Shen, Distinct Fermi-momentum-dependent energy gaps in deeply underdoped Bi2212, Science **314**, 1910 (2006).

[39] H. L. Edwards, J. T. Markert, and A. L. de Lozanne, Energy gap and surface structure of $YBa_2Cu_3O_{7-x}$ probed by scanning tunneling microscopy, Phys. Rev. Lett. 69, 2967 (1992).

[40] V. Y. Butko, G. Logvenov, N. Božović, Z. Radović, and I. Božović, Madelung strain in cuprate superconductors – a route to enhancement of the critical temperature, Adv. Mater. **21**, 3644-3688 (2009).

[41] N. Driza, S. Blanco-Canosa, M. Bakr, S. Soltan, M. Khalid, L. Mustafa, K. Kawashima, G. Christiani, H-U. Habermeier, G. Khaliullin, C. Ulrich, M. Le Tacon, and B. Keimer, Long-range transfer of electron-phonon coupling in oxide superlattices, Nature Mater. **11**, 675-681 (2012).

[42] H. Zhou, Y. Yacoby, V. Y. Butko, G. Logvenov, I. Božović, and R. Pindak, Anomalous expansion of the copper-apical oxygen distance in superconducting $La_2CuO_4$−$La_{1.55}Sr_{0.45}CuO_4$ bilayers, Proc. Nat. Acad. Sci. **107**, 8103-8107 (2010).

[43] Y. Yu, L. Ma, P. Cai, R. Zhong, C. Ye, J. Shen, G. D. Gu, X. H. Chen, and Y. Zhang, High-temperature superconductivity in monolayer $Bi_2Sr_2CaCu_2O_{8+\delta}$, Nature **575**, 156-163 (2019).